# Multiple images storage and frequency conversion in a cold atomic ensemble

Dong-Sheng Ding[†], Jing-Hui Wu, Zhi-Yuan Zhou, Bao-Sen Shi[*], Xu-Bo Zou, and Guang-Can Guo

*Key Laboratory of Quantum Information, University of Science and Technology of China, Hefei 230026, China*

*Corresponding author:* [*]drshi@ustc.edu.cn

[†]dds@mail.ustc.edu.cn

**Abstract**

The strong demand for quantum memory, a key building block of quantum network, has inspired new methodologies and led to experimental progress for quantum storage. The use of quantum memory for spatial multimode or image storage could dramatically increase the channel bit-rate. Furthermore, quantum memory that can store multiple optical modes would lead to higher efficiencies in quantum communication and computation. Here, by using resonant tripod electromagnetically induced transparency in a cold atomic ensemble, we experimentally demonstrate multiple probes storage in frequency domain, where two probe fields have discrete wavelengths and different spatial information. In addition, by using different read-light, we realize frequency conversion of retrieved images with high efficiency. Besides, our method could be used to create a superposition of the images by realizing the function of a beamsplitter. All advantages make our method useful in many fields including quantum information, detection, imaging, sensing and even astrophysical observation.

**Introduction**

A long-distance quantum communication network consists of a memory in which quantum information can be stored and manipulated at will and a carrier via which the different memories can be connected for. The seminal work of Duan *et. al*. [1] shows that an atomic system could be a suitable candidate for the memory and a photon could be a robust and efficient carrier due to its weak interaction with the environment.

The strong demand for quantum memory has inspired new methodologies and led to experimental progress for quantum storage using an atomic system via different mechanisms, for example, via electromagnetically induced transparency (EIT) [2, 3], atomic frequency combs [4, 5], Raman schemes [6, 7], and gradient echo memory [8, 9], etc. Even so, there has been little progress on spatial multimode or image storage.

A quantum memory which could store spatial multimode or image is very useful because it allows for simultaneous storage of multiple signals in a single storage device. It could dramatically increase the channel bit-rate. Recently there have been some progresses related to the storage of images, for example, via EIT in a hot atomic ensemble or a cryogenically cooled doped solid, using the four-wave mixing technique in a hot or cold atomic ensemble and via a gradient echo memory in a hot atomic ensemble [15]. Additionally, quantum memories that are able to store multiple optical modes offer advantages over single-mode memories in terms of speed and robustness, leading to higher efficiencies in quantum communication and computation experiments [16, 17]. There have been some experimental reports in this direction, using the gradient echo memory in a hot atomic ensemble [8] or using the technique of the spectral shaping of an inhomogeneously broadened optical transition into an atomic frequency comb in solids doped with rare earth metal ions [4]. Very recently, we reported the experimental demonstration in the spatial domain through EIT in a cold $^{85}$Rb atomic cloud [18]. So far, there had been no reports on the storage of multimode images in frequency domains.

A promising way of storing multiple optical modes is to store the different optical modes in different atomic collective spin excitation states, which could, for example, be realized by a tripod configuration of atoms. There have been some related progresses (no image) along this direction of research [19, 20]. In this paper, we report the first experimental evidence that two different images can be stored in the frequency domain using EIT in a tripod configuration in a cold atomic ensemble. The development of multimode quantum image memories should be an important step towards the realization of high-dimensional quantum networks. Furthermore, by using different read-light, we realize the highly-efficient frequency conversion of the stored images. The high conversion efficiency is achieved even the input image with about $10^4$ photons per pulse and very weak control power (about 100 μW) are applied. It is about 3 orders higher than that achieved in a recent upconversion experiment by the four-wave mixing in a hot atomic ensemble [21, 22], is much higher than that (about 13%) obtained by intracavity four-wave mixing in nonlinear Fabry-Perot interferometer [23] and is also higher that that (40%) obtained through a three-wave mixing process in a cavity by using a high power solid laser [24]. Although the wavelength difference between the input image and the converted image is only about 15 nm in our experiment, the frequency conversion of images between different wavelength bands could be realized by selecting suitable atomic energy levels or a suitable atom. The image conversion could be realized in a hot atomic ensemble too, therefore the whole experimental system could be much simplified and miniaturized. We believe that such a technique would may find wide applications in detection, imaging, sensing and even astrophysical observation [25-28]. Besides, we could create the superposition of the two images by realizing the function of a beamsplitter, which is a key element of quantum information processing. The big difference in this respect compared with previous reported work [29] is that we could create the superposition of images.

We consider the state of input probe fields being expressed below:

$$|\psi_{input}\rangle = |\lambda_1\rangle|I_1\rangle + |\lambda_1'\rangle|I_2\rangle , \qquad (1)$$

where, $\lambda_{1,2}$ indicates wavelength, $I_{1,2}$ describes spatial information, $|\lambda_1\rangle|I_1\rangle$ and $|\lambda_1'\rangle|I_2\rangle$ describe input probes 1 and 2 with wavelengths $\lambda_1$ and $\lambda_1'$, respectively. The frequency shift between these two probe fields is defined as $\Delta$. If we use the $\lambda_1$ field as read-light, then the retrieved probe 1, 2 fields are expressed as:

$$|\psi_{output}\rangle = \sqrt{\eta_1}|\lambda_1\rangle|I_1\rangle + \sqrt{\eta_2}|\lambda_1'\rangle|I_2\rangle , \qquad (2)$$

where, $\sqrt{\eta_1}|\lambda_1\rangle|I_1\rangle$ describes the retrieved probe 1 at $\lambda_1$ with storage efficiency of $\eta_1$, $\sqrt{\eta_2}|\lambda_1'\rangle|I_2\rangle$ corresponds to the retrieved probe 2 at $\lambda_1'$ with storage efficiency of $\eta_2$. If we apply a $\lambda_2$ field as read-light, then the readout probe fields can be described by:

$$|\psi_{output}\rangle = \sqrt{\eta_3}|\lambda_2\rangle|I_1\rangle + \sqrt{\eta_4}|\lambda_2'\rangle|I_2\rangle , \qquad (3)$$

where, $\sqrt{\eta_3}|\lambda_2\rangle|I_1\rangle$ describes the retrieved probe 1 at $\lambda_2$ with conversion efficiency of $\eta_3$, and $\sqrt{\eta_4}|\lambda_2'\rangle|I_2\rangle$ corresponds to the retrieved probe 2 at $\lambda_2'$ with conversion efficiency of $\eta_4$. The frequency shift between $\lambda_2$ and $\lambda_2'$ is still $\Delta$. The frequencies of retrieved probes are changed in comparison with Eq. (2), therefore the frequency conversion of the stored image could be realized.

**Experimental configuration**

Figure 1 shows the schematic experimental setup. A tripod configuration shown in Fig. 1(b) was used to perform the storage experiment. A cigar-shaped atomic cloud of $^{85}$Rb atoms, trapped in a two-dimensional magneto-optical trap (MOT), was used as the storing media. The size of cloud is about $30\times2\times2$ mm$^3$. The total atom number is $9.1\times10^8$.[30] The probe fields (probe 1 and 2) and the write (W)/read (R) beam from an external-cavity diode laser (ECDL, DL100, Toptica) have the same wavelength of 795-nm. The probe fields were imprinted a real image through a standard resolution chart (USAF target). The other read (R') beam from another ECDL has the wavelength of 780-nm. Probe 1 ($P_1$) and the W(R) fields are resonant with the transitions $5S_{1/2}$ (F=3, $m_F$=-3) ($|2\rangle$)–$5P_{1/2}$ (F'=2, $m_F$=-2) ($|4\rangle$) and $5S_{1/2}$ (F=3, $m_F$=-1) ($|1\rangle$)–$5P_{1/2}$ (F'=2, $m_F$=-2) ($|4\rangle$), respectively. Probe 2 ($P_2$) is resonant with the transition $5S_{1/2}$ (F=2, $m_F$=-2) ($|3\rangle$)–$5P_{1/2}$ (F'=2, $m_F$=-2) ($|4\rangle$). The R' field is resonant with the transition $5S_{1/2}$ (F=3, $m_F$=-1) ($|1\rangle$)–$5P_{3/2}$ (F'=3, $m_F$=-2) ($|5\rangle$). $P_1$ and $P_2$ co-propagate with a small angle of 0.1°. The angle between $P_1$ and the read beam is about 2.5°. The non-collinear configuration used in the experiment significantly reduces the noise from the scattering of the write/read light. We used two photomultiplier tubes (PMT) (Hamamatsu, H10721) to detect the intensities of probe fields in the time domain and used a time-resolution camera (CCD, 1024×1024, iStar 334T series, Andor) to monitor spatial structure. By adjusting a quarter-wave plate before the MOT, the write/read and probe fields were assigned opposite circular polarizations. Using a quarter-wave plate after the MOT, the fields were later reversed to have orthogonal linear polarizations.

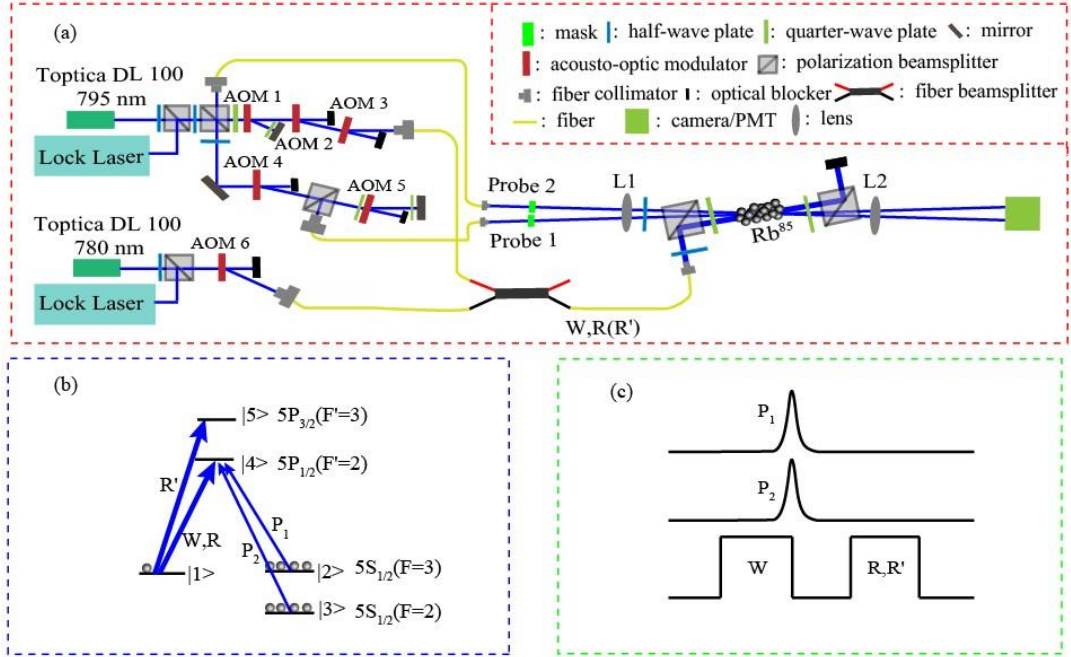

**Figure 1** (a) Experimental setup. The frequency of probe 1 ($P_1$) is shifted by AOM 4 (frequency shift 160 MHz) and AOM 5 (double-pass frequency shift -160 MHz); The frequency of probe 2 ($P_2$) is shifted by AOM 1 (double-pass frequency shift 3.0378 GHz); The frequency of W(R) is shifted by AOM 2 (frequency shift 80 MHz) and AOM 3 (frequency shift -80 MHz); The read beam R' is modulated by AOM 6. (b) Experimental energy diagram. (c) Timing sequence. Probe fields are modulated by acoustic-optical modulators to form a Gaussian pulse sequence. The W field writes the probe fields into the atomic collective spin excitation states and the R, R' fields read them out.

**Experimental results**

We take the experimental parameters to be as follows: $\lambda_1$=795 nm, $\lambda_1'$=795 nm (the frequency shift between them is $\Delta$= 3.0378 GHz); $I_1$ and $I_2$ describe the spatial information of the digit "2" and the image " " respectively. The wavelength of R is $\lambda_1$=795 nm. The W(R) field and R' field are 3 mm in diameters and cover the probe beams completely. If we use the R field as read-light, the retrieved signals have a wavelength of 795nm. The signals obtained by PMTs are shown in figure 2(a). The storage efficiency of probe 1 is $\eta_1=0.098$, the storage efficiency of probe 2 is $\eta_2=0.023$. Then, we change the PMTs to CCD for detecting spatial information. The retrieved images are shown in figure 2(b). The left column corresponds to the storage of probe 1 and the middle column is the storage of probe 2. The case of probe 1 and probe 2 being simultaneously stored and retrieved is shown in the right column of Figure 2(b). There is almost no difference between the leakage images and the retrieved images. We experimentally demonstrate that there is no crosstalk between the storage of probe 1 and probe 2. The detailed experimental results are shown in Fig. 1-3 of supplementary material.

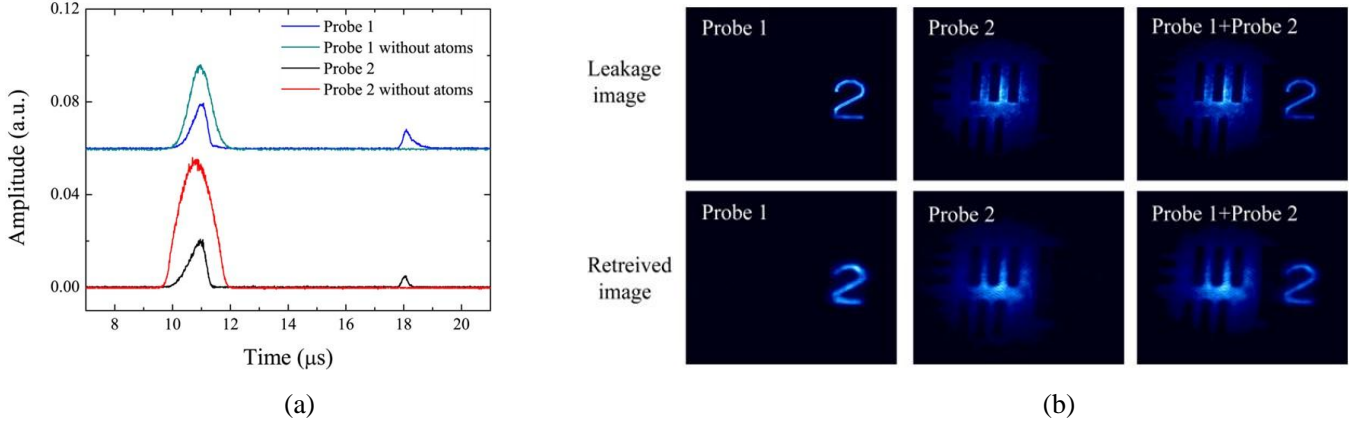

(a)            (b)

**Figure 2** (a) The intensities of the leakage and retrieved probe fields recorded by two PMTs. (b) The leakage and retrieved images recorded by CCD. Each image is the sum of the 50 retrieved images. The exposure time of the CCD camera was 1.0 s. The probes 1 and 2 are stored for about 6.7μs time. The power of the W(R) field is 100μW and the R' field is 140μW. The photon number of each probe is about $2.7\times 10^4$

Next, we want to see what will happen if a different read-light R' with different wavelength is used. In order for that to occur, we use the R' field with the wavelength of $\lambda_2=780$ nm as the read-light, the results in time domain is shown in figure 3(a). The leakage signals have a wavelength of 795nm, but the retrieved signals have a wavelength of 780nm. The frequency conversion efficiency of probe 1 is $\eta_3=0.017$, the efficiency of probe 2 is $\eta_4=0.018$. Figure 3(b) is a record of the spatial structures of retrieved probes using CCD. The experimental results show that the frequency of the output fields can be changed using different read-light and the spatial information of the probe fields could still be preserved. Therefore we realize the image storage and its frequency conversion simultaneously in a single storage device. We also find that the efficiency of conversion decreases with the increase of the angle between the probes and the write/read light. The detail experimental results are shown in Figs. 4 and 5 of supplementary material.

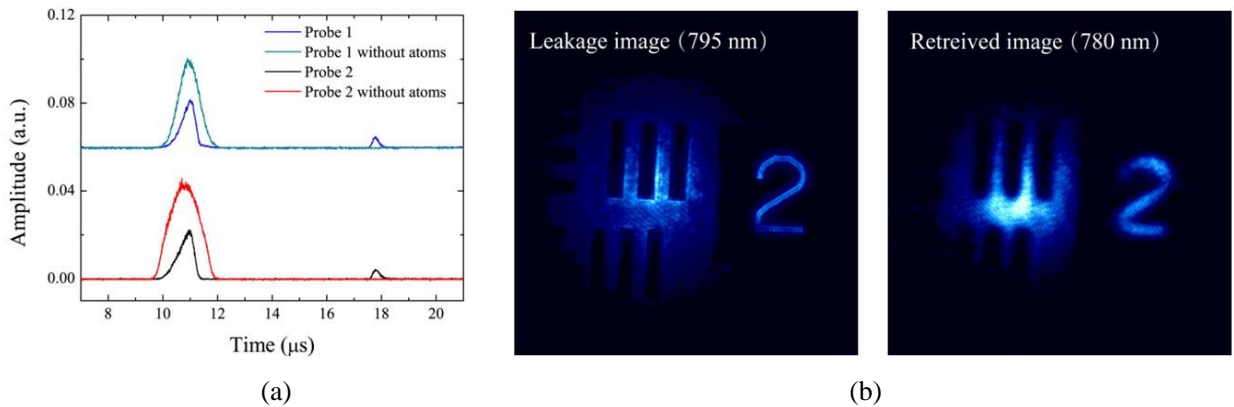

(a)            (b)

**Figure 3** (a) is the intensity signal in the time domain recorded by PMTs. The storage time of probe fields is about 6.7 μs. The left column of (b) is the leakage image with a wavelength of 795 nm, right column is the retrieved image with a wavelength of 780 nm. Each image is the sum of the 50 retrieved images. The exposure time of the CCD camera was 1.0s. The image storage time is about 6.7 μs time. The power of the W(R) field is 100 μW and the R' field is 140 μW. The photon number of each probe is about $2.7 \times 10^4$.

The experiment shown before has clearly demonstrated that a tripod configuration of atoms could be used to realize frequency conversion of images. For such as a process, the frequency conversion efficiency between the frequency-converted image and the input image is an important parameter for evaluating its workability. In the following, we check the frequency conversion efficiency. We only consider the conversion of probe 1 imprinted with an image of three-slit structure from 795nm to 780nm. The experimental result is shown in Fig. 4. The efficiencies of storing probe 1 with R read and R' read light are 0.63 and 0.55. The visibilities of retrieved 795 nm and 780 nm images are 80% and 74%. The big difference in conversion efficiencies between the experiment shown in Fig. 4 and the experiment shown in Fig. 3 using the tripod system is mainly caused by the follow reasons: one is from different the populations of the ground states. In the case of Fig. 3, the repumping field of the MOT has to be turned off during the storage experiment, the atoms of energy levels |2> are pumped to energy levels |3> when W field is applied. On the contrary, in the case of Fig. 4, the repumping light is always on, so almost all atoms are pumped to level |2>, therefore a more large optical depth is achieved in this case, which induces the much smaller group velocity with the probe 1 propagates in the cloud. The smaller group velocity in the case of Fig. 4 makes the storage more easily and efficiently. Another reason is from the angle between the probe and W(R) light, as shown in Fig. 5 of supplementary material. The smaller the angle is, the higher the efficiency is. The angle in case of Fig. 3 is larger than that in Fig. 4. Besides, another possible reason is from experimental setup. Optimizing the setup in the case of Fig. 3 is much more difficult than that in the case of Fig. 4. In addition, the significant population in level |2> may make the probe 1 in the case of Fig. 4 experience gain on retrieval [31], which causes the high output.

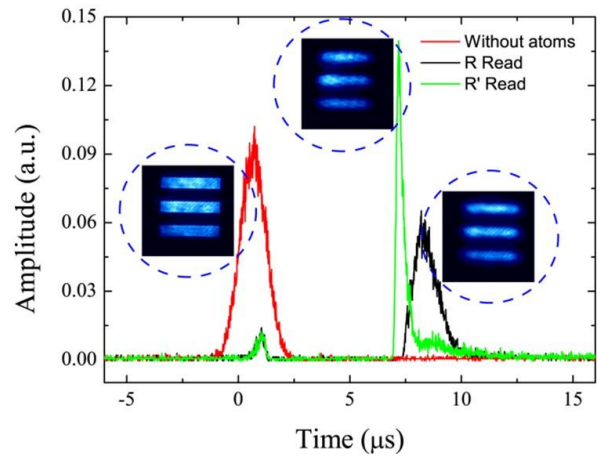

**Figure 4** The write fields are both W fields. The angle of probe 1 and coupling fields is 1.2 °. The power of the W(R) field is 100μW and the R' field is 140μW. The photon number of probe is about $2.7 \times 10^4$

A potential interesting thing is if two images are stored overlapping, what will happen? Does the retrieved light contain an interesting pattern? Or is the intensity simply added? Our experimental results answer these questions and clearly demonstrate that the storage is coherent. The details are shown in supplement materials.

**Discussion and analysis**

The retrieved images are very similar to the input images, even when a different read-light are used. This illustrates that the spatial properties $|I_j\rangle$ in state $|\lambda_n\rangle|I_j\rangle$ do not change during the process of storing and retrieving, and the process of frequency conversion. Moreover, the readout images between the probe fields have no crosstalk between them. In figure 3, there are about $2.7 \times 10^4$ photons per pulse being stored, which is higher than the number of photons in Ref. [18]. In this experiment, the atomic populations of energy levels |2> in this tripod configuration are relatively small compared with the degenerate configuration in Ref. [18]. Then, the storage efficiency is lower compared with the degenerate configuration. Therefore, the repetition of storage experiment per second is hard to increase. In figure 3 and 4, we observe that the input state $|\psi_{input}\rangle = \sum |\lambda_i\rangle|I_j\rangle$ is converted into the state $|\psi_{output}\rangle = \sum \sqrt{\eta_n} |\lambda_n\rangle|I_j\rangle$, where $(n,i,j)= (1, 2)$. This process can be used to realize the function of a beamsplitter, by which the superposition of the different images at different wavelengths is obtained. This function can be realized by storing only part of the probes, such that the leakage image and the retrieved image consist of the superposition [29]. We could realize two different superposition images states: one is superposition state of the leaked part and the retrieved part without frequency conversion, another is the superposition state consisted of the leaked part with wavelength of $\lambda_1$ and the retrieved part with the wavelength of $\lambda_2$ when the frequency conversion is applied. It is worth illustrating that efficiency of the frequency conversion is much higher than difference frequency generation using a nonlinear crystal or a four-wave mixing inan atomic system. The frequency conversion is essentially four-wave mixing, a process which is split into two delayed parts, with a long-lived coherence in the atoms connecting the storage and retrieval processes. The atoms are coherently prepared during the process of storage. It has been demonstrated previously that coherently prepared media may enhance nonlinear optical interaction [31, 32]. Therefore the efficiency is much high compared with other works. Although the wavelength difference between the input image and the converted image is only about 15 nm in our experiment, we could realize the frequency conversion of images between different wavelength band by selecting suitable atomic energy levels or selecting a suitable atom. For example, if the energy level of 6p of the $^{85}$Rb atom is considered, then the image could be transferred from near infrared band (795 nm) to ultraviolet band (420 nm) or reversely. Although our experiment was done in a cold atomic ensemble, it could be realized in a hot atomic ensemble, therefore the whole experimental system could be much simplified and miniaturized.

In summary, we have experimentally demonstrated that multiple images can be stored and retrieved in frequency domain in a cold atomic ensemble using resonant tripod EIT configuration. The probe fields can be converted to different frequencies by applying different read-light while the spatial information is preserved. Our results are important for future quantum communication in high-dimensional quantum networks and other fields.

Note: during the preparation of this manuscript, we noticed a parallel work [32] appeared in arxiv:1205.1495 (2012). In this work, the gradient photon echo technique was used to realize the storage of multiplex images in a hot atomic ensemble.

**Acknowledgments**

This work was supported by the National Natural Science Foundation of China (Grant Nos. 10874171, 11174271, 61275115), the National Fundamental Research Program of China (Grant No. 2011CB00200), and the Innovation fund from CAS, Program for NCET.

# Supplementary material

# Multiple images storage and frequency conversion in a cold atomic ensemble

**Dong-Sheng Ding[†], Jing-Hui Wu, Zhi-Yuan Zhou, Bao-Sen Shi[*], Xu-Bo Zou, and Guang-Can Guo**
*Key Laboratory of Quantum Information, University of Science and Technology of China, Hefei 230026, China*

Corresponding author: [*]*drshi@ustc.edu.cn*
[†]*dds@mail.ustc.edu.cn*

**Method**

We used a resonant tripod EIT configuration shown in figure 1(b) in text. A cigar-shaped atomic cloud of $^{85}$Rb atoms was obtained in a two-dimensional magneto-optical trap (MOT) and served as the memory element in our experiment [1]. The size of cloud is about $30\times2\times2$ mm$^3$. The total atom number is $9.1\times10^8$. The non-degenerate states |2> and |3> were used in our system. We used a 1.5 GHz AOM to generate probe 2 with double-pass light and a frequency shift of 3.0378 GHz relative to probe 1 or the R field. Input signals were modulated into Gaussian profiles by an arbitrary signal generator (AFG 3252). A high time-resolution camera (1024×1024, iStar 334T series, Andor) was used to monitor the spatial structure of the probe fields, including a sequence of leakage pulses and retrieved pulses. The camera has a quantum efficiency of approximately 25% at 795 and 780nm. The camera with its high speed shutter could be triggered by an external TTL signal. This TTL signal was generated from AFG 3252 and delayed by a delay generator (DG 535). The total size of the camera sensor was 13.3×13.3mm. Our geometry was based on a 4-f imaging system, consisting of the center plane of the atomic cloud, the imaging plane of the camera, the mask plane, and the two lenses. L1 and L2 are lenses with a focal length of 300mm and 500mm, respectively. In our experiment, the full input pulse was divided equally into two parts. One part was stored and the other part was leaked. The probes with images are collected by a multimode fiber connected to PMTs. The efficiency is defined as the comparison of the area under the curve between the retrieved signal to the input signal without atom.

## 1. Cross-talk verification

In this section, we try to answer the question: whether there is cross-talk between the stored images. We provide the experimental evidences in three different cases: in case 1, the probe 1 and probe 2 have same frequencies. We separately consider the storage of probe 1, probe 2 and probe 1+2. The experimental results are shown in Fig. 1, where, (a),(c),(e) are the leaked images of probe 1, probe 2 and probe 1+2, and (b), (c), (d) are retrievals of them. The results clearly show that there is no cross-talk between two stored images.

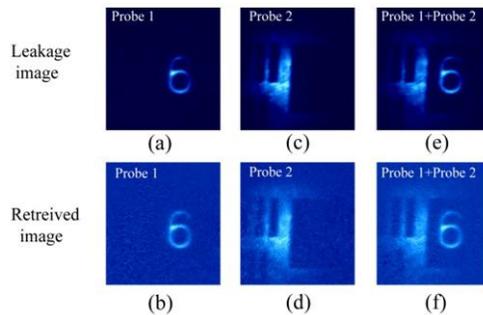

Fig. 1 The storages of two probe fields with the same frequencies. (a) and (b) are leaked and retrieved images of probe 1. (c) and (d) are for the probe 2 and (e) and (f) for the probe 1+2.

In case 2, we consider the crosstalk against the angle between the write/read and probe fields, where the probes have different frequencies. The angle between the write/read and probe 1 is defined as α; the angle between the probe 1 and probe 2 is defined as β. The experimental results are shown in Fig. 2. The left column in the each picture is for probe 1; the middle column is for probe 2 and the right column is for probe 1+2. Each picture is the extract from the background. The upper column is leaked image and the bottom of the picture corresponds to the retrieval. In this process, the storage time is about 7 μs. In Fig. 2, the angle β is kept to be $0.5°$. Obviously, there is also no cross-talk in this situation.

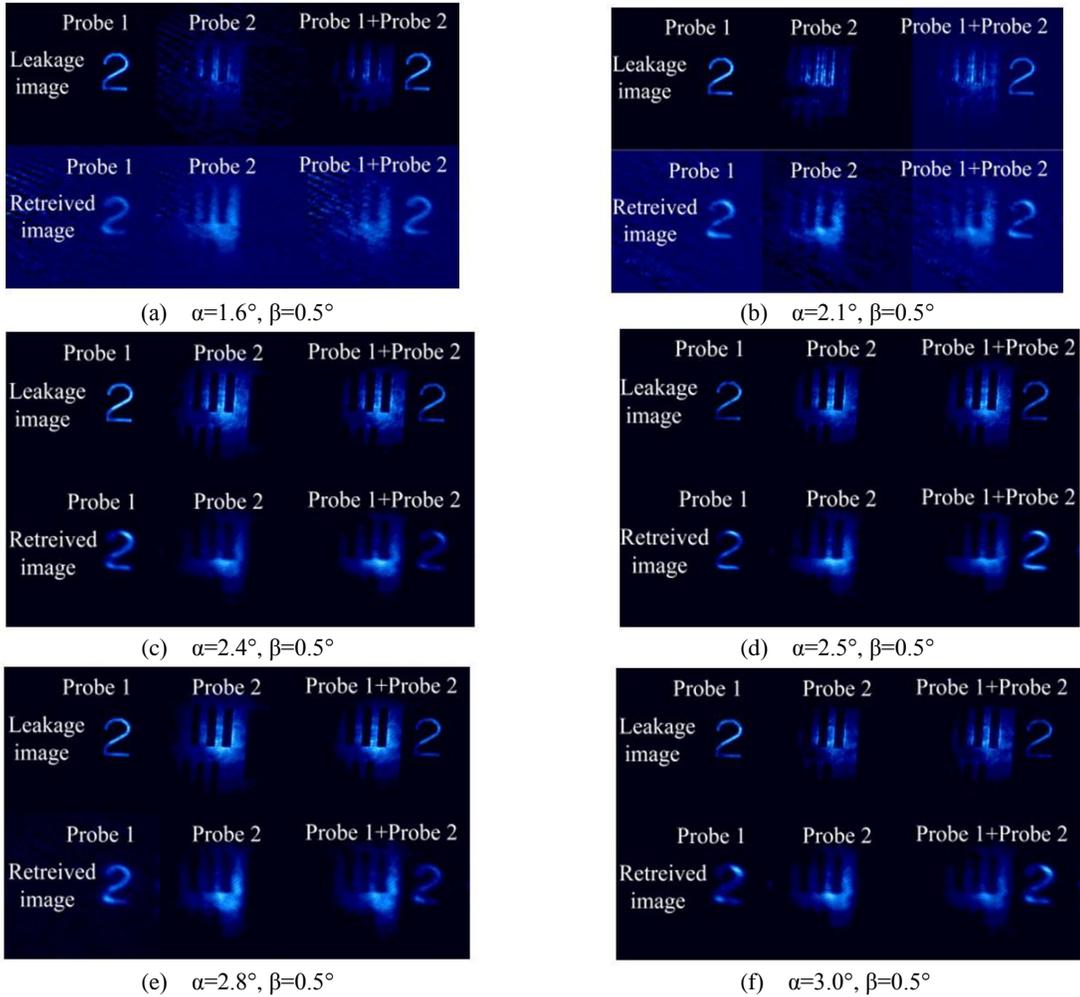

(a) α=1.6°, β=0.5°  (b) α=2.1°, β=0.5°
(c) α=2.4°, β=0.5°  (d) α=2.5°, β=0.5°
(e) α=2.8°, β=0.5°  (f) α=3.0°, β=0.5°

Fig. 2 The crosstalk between the images against the angle between α. The angle β is kept to be β=0.5° in all measurements. When α is small, the noise from W/R field is difficult to reduce, so figure. (a) and (b) have little blur.

In case 3, we keep the angle of α being 3.5° unchanged, and check the cross-talk against the change of the angle β. In our experiment, when the angle is larger than 1.33°, the camera can't collect both probes simultaneously. If the angle β is too small, the two images are overlapped. therefore, we only change the angle between β=0.5°~β=1.33°. In this process, the storage time is about 7 μs. The experimental results are shown in Fig. 3. It clearly shows that there is no cross-talk between the stored images.

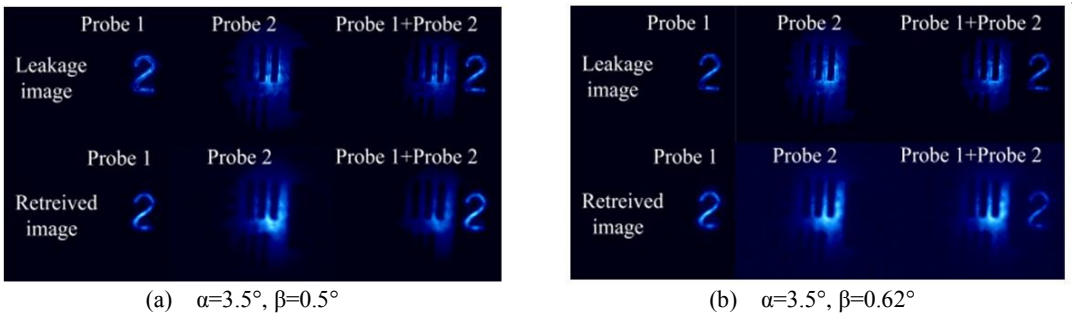

(a) α=3.5°, β=0.5°  (b) α=3.5°, β=0.62°

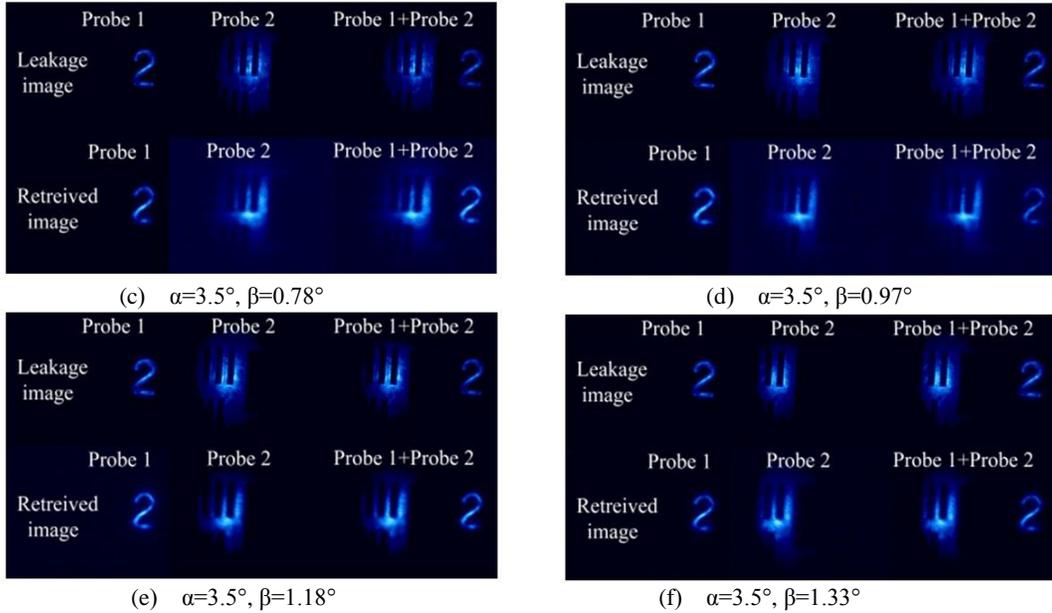

(c) α=3.5°, β=0.78°  (d) α=3.5°, β=0.97°

(e) α=3.5°, β=1.18°  (f) α=3.5°, β=1.33°

Fig. 3 The crosstalk between the images against the angle of β. The angle of α is kept to be 3.5 ° in all measurements.

In conclusion. There is no cross-talk between the stored images.

## 2. Efficiency

In this section, we check the efficiencies of the storages against the angles α and β. In this process, we only consider the storage of the probe 1. Besides, because what we want to learn is how these angles affect the efficiency, therefore we only consider the storage of probe 1 without image for simplification. We believe that the results obtained could be applied to the case with image. The efficiencies of storage of probe 1 with R read light and R' read light are defined as $\eta_1$, $\eta_2$. The storage time is 6 μs. In this process, the powers of R and R' fields are 106 μW and 320 μW. The experimental results are shown in Fig. 4.

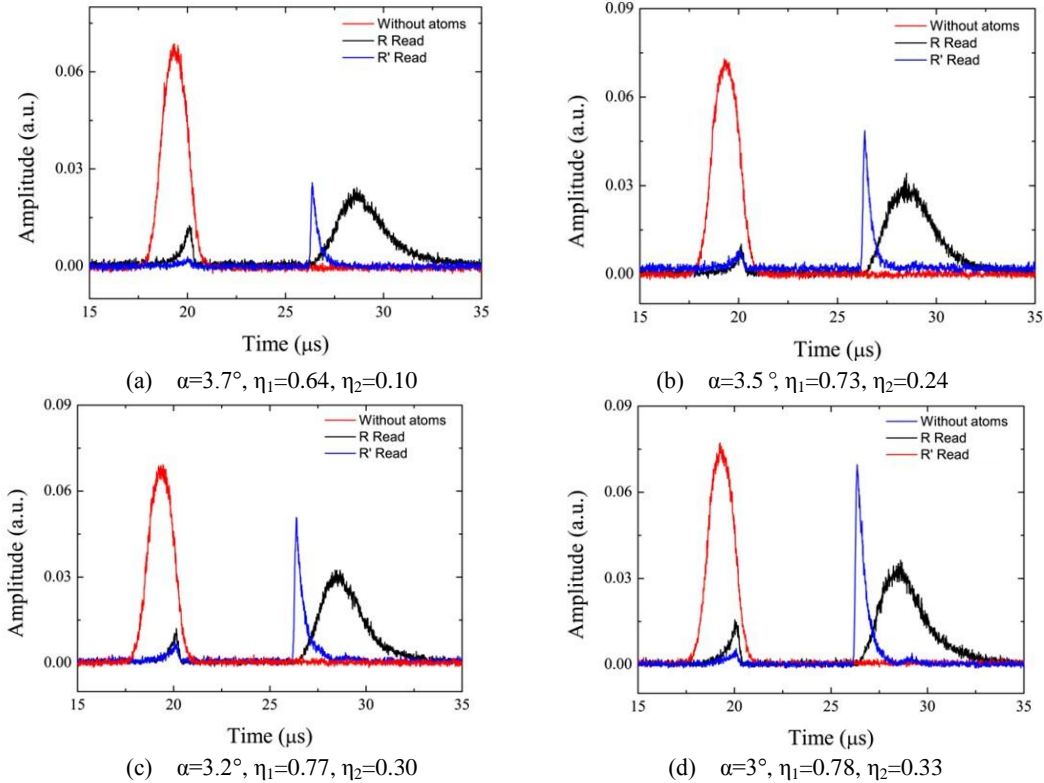

(a) α=3.7°, $\eta_1$=0.64, $\eta_2$=0.10  (b) α=3.5°, $\eta_1$=0.73, $\eta_2$=0.24

(c) α=3.2°, $\eta_1$=0.77, $\eta_2$=0.30  (d) α=3°, $\eta_1$=0.78, $\eta_2$=0.33

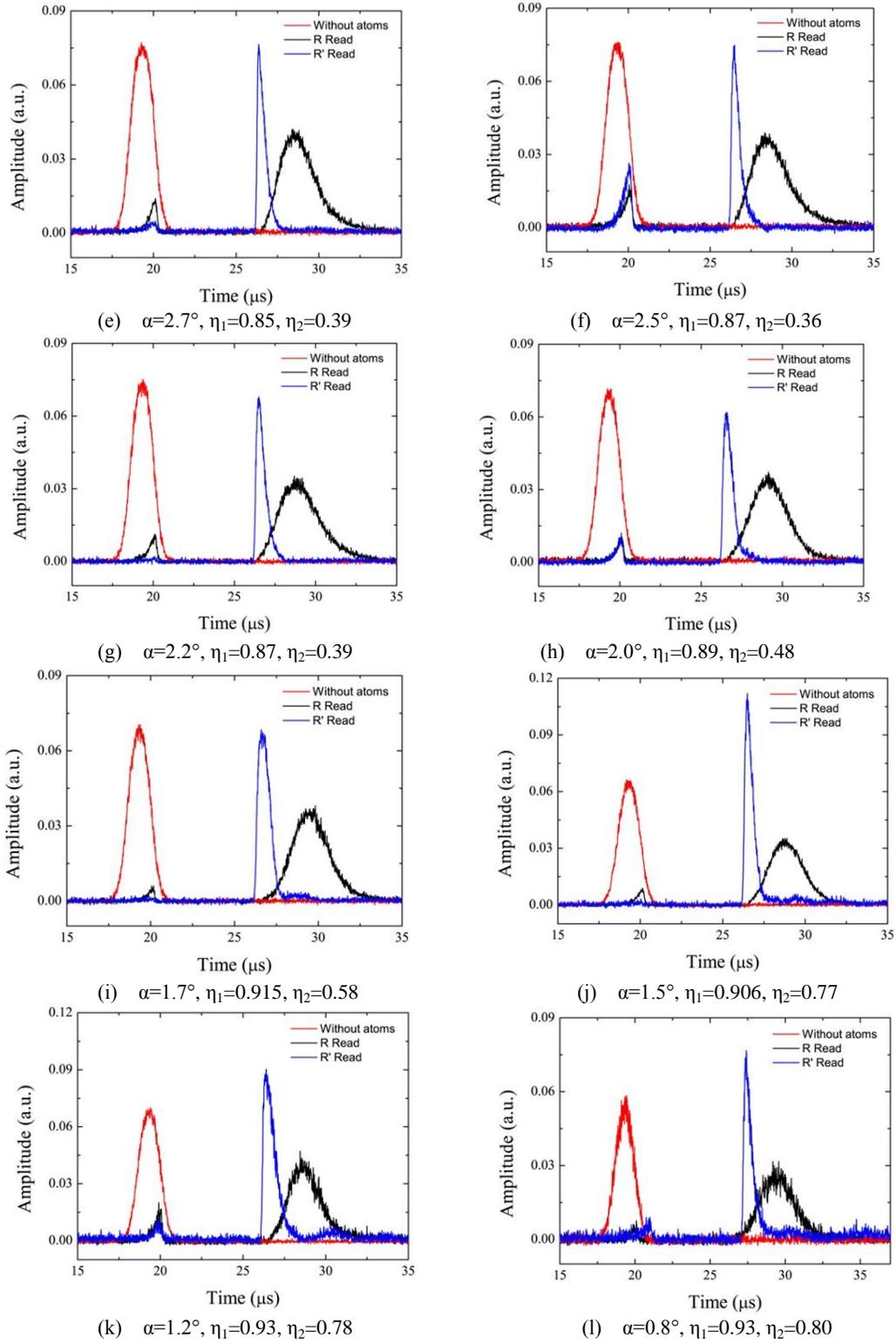

(e) α=2.7°, η₁=0.85, η₂=0.39
(f) α=2.5°, η₁=0.87, η₂=0.36
(g) α=2.2°, η₁=0.87, η₂=0.39
(h) α=2.0°, η₁=0.89, η₂=0.48
(i) α=1.7°, η₁=0.915, η₂=0.58
(j) α=1.5°, η₁=0.906, η₂=0.77
(k) α=1.2°, η₁=0.93, η₂=0.78
(l) α=0.8°, η₁=0.93, η₂=0.80

Fig. 4 The stored efficiencies of probe 1 using different read light R and R' against angle of α.

We then draw the efficiencies of storage of probe 1 against the angle of α, which is shown in Fig. 5. The experimental results clearly show that the efficiency decreases with the increase of the angle.

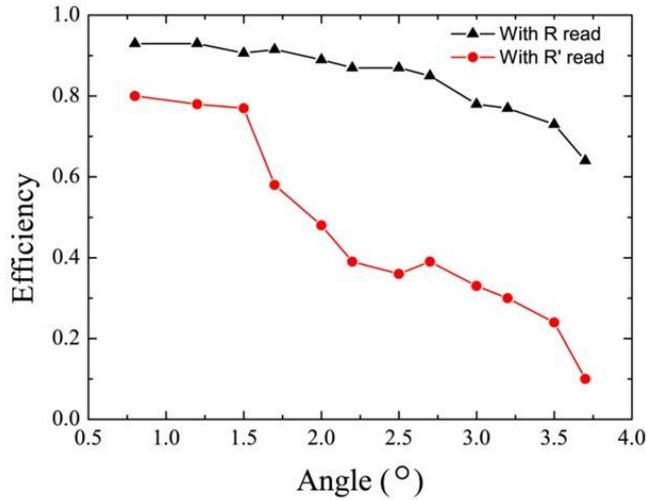

Fig. 5 The efficiencies of storage vs angle α.

## 3. Demonstration of the coherence

A potential interesting thing is if two images are stored overlapping, what will happen? Does the retrieved light contain an interesting pattern? Or is the intensity simply added? We perform the storage experiments in two different conditions to answer these questions. In order to simplify experiment, we only consider the storages of probes without images. In case 1, the probe 1 and 2 have different frequencies. The probe 1 is resonant on the transition $5S_{1/2}$ (F=3) (|2>)–$5P_{1/2}$ (F'=2) (|4>), the probe 2 is resonant on the transition $5S_{1/2}$ (F=2) (|3>)–$5P_{1/2}$ (F'=2) (|4>). In case 2, the probe 1 and probe 2 have same frequencies and are all resonant on the transition $5S_{1/2}$ (F=3) (|2>)–$5P_{1/2}$ (F'=2) (|4>). In these two cases, R read (795 nm laser) light is used to read the stored probe 1 and probe 2 out. The results are shown in figure 6. The retrieved interference phenomenon is only observed on the condition that the probe 1 and probe 2 have same frequencies. In case 1, we could not observe the interference pattern due to the large frequency difference (3GHz) between probes. The results we obtained clearly demonstrate that the storage is coherent.

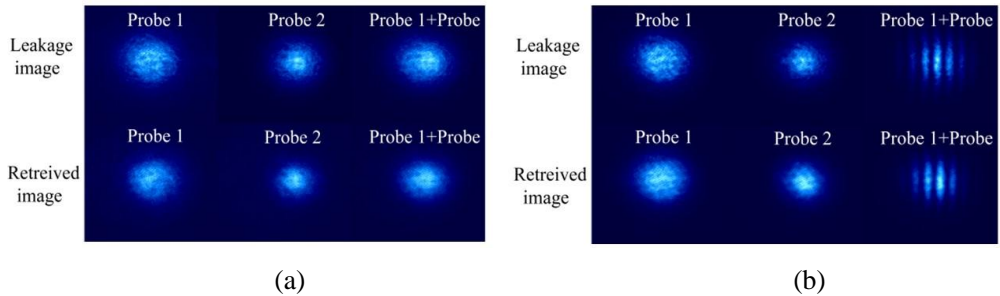

(a)                                   (b)

Fig. 6 The storage under the condition of overlapping of probe 1 and probe 2. (a) The probe 1 and probe 2 have 3.0 GHz frequency difference. (b) The probe 1 and probe 2 have same frequencies.

## 4. Storage near single-photon level

In this section, we consider the workability of the scheme at near single-photon level. In the experiment, by using an attenuator, we reduced the intensity of the probe fields and detected the leakage and retrieved images. There are about $1.3 \times 10^3$ photons per pulse being stored. We obtained the results shown in Fig. 7; the retrieved images are very similar to the leaked images. We believe that the scheme might work at the lower power of coherent light if we increase the density of the atoms and the repetition rate of the probes further. Very recently, a single-photon level image storage experiment has been done in our lab. In that experiment, a high repetition rate for probe fields and the coupling field is employed, which leads to a high signal-to-noise ratio (SNR) at the single-photon level.

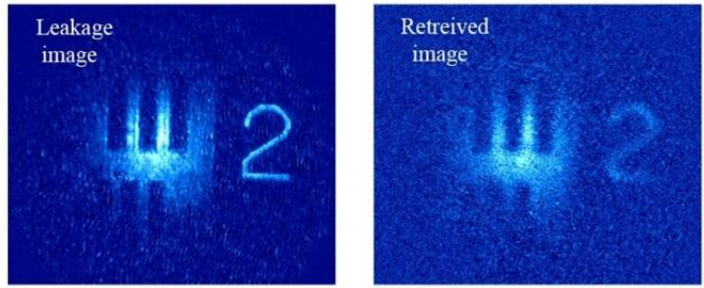

Fig. 6 Leakage and the retrieved images using the R field as read-light near single-photon level. The left column is the leakage image; the right column is the retrieved image. Each image is the sum of the 200 retrieved images. The exposure time of the CCD camera was 1.0 s.